# Distinguishing AI-Generated Music from Edited Audio as a Hard-Negative Robustness Task

**Alexandru-Ştefan Moroşanu**[1] , **Valerian Cecan**[1] , **Ştefan-Daniel Achirei**[1] , **Laura Erhan**[1,2]
[1]Department of Computer Science and Engineering, Faculty of Automatic Control and Computer Engineering, Gheorghe Asachi Technical University of Iaşi, Iaşi, Romania
[2]Free University of Bozen-Bolzano, Bolzano, Italy
{alexandru-stefan.morosanu, valerian.cecan}@student.tuiasi.ro, {stefan-daniel.achirei, laura.erhan}@academic.tuiasi.ro {laura.erhan}@unibz.it

## Abstract

AI-generated music detectors are commonly evaluated against original songs, but real-world uploads are often remixed, re-encoded, pitch-shifted, or otherwise edited. These edited versions form a difficult negative class: they are not generated by AI, yet they may introduce spectral artifacts that resemble synthetic audio fingerprints. We study this problem as a hard-negative robustness setting for AI-generated music detection, focusing on AI-generated and edited variants derived from the same anchor songs. We compile a YouTube-based dataset of AI, edited, and original variants, using the original tracks only as references, and train a binary AI-versus-edited detector. Audio is processed as 10-second clips and passed as raw waveforms to a pretrained PaSST spectrogram transformer. To reduce leakage, all splits are performed by anchor song. On the held-out test set, the final video-level system achieves 0.811 balanced accuracy. At clip level, AI-generated clips reach an F1-score of 0.836, while edited clips reach a lower F1-score of 0.720. The results suggest that AI-generated music retains detectable fingerprint-like spectral cues beyond ordinary editing, but the lower edited-class F1-score shows that these cues can still overlap with artifacts from edited audio. Grad-CAM visualizations are used to inspect whether high-confidence predictions rely on localized time–frequency regions.

## 1 Introduction

Music generation systems have moved from short experimental demos they have become full-song production tools. Current models can generate complete tracks with vocals, instrumentation, arrangement, and production style from short text prompts or minimal user input. Systems such as MusicGen show that transformer-based audio token models can generate coherent music from text and melody conditioning [Copet *et al.*, 2023], and commercial platforms such as Suno and Udio make this type of generation widely accessible. This creates a practical need for automatic methods that can identify AI-generated music in online catalogues, streaming platforms, and user-generated content repositories.

Recent work shows that AI-generated music can often be detected from audio alone. Large-scale studies on Suno and Udio tracks report that classifiers trained on audio embeddings can separate human-made and AI-generated music with strong performance, but also show that the behavior of the detector can be sensitive to transformations such as resampling and other signal changes [Cros Vila *et al.*, 2025]. Other studies associate detection cues with artifacts introduced by the generative audio pipeline itself. [Afchar *et al.*, 2025a] argue that synthetic music may contain systematic spectral artifacts associated with decoder and deconvolution operations, and use this observation to motivate interpretable detection criteria. These findings suggest that AI-generated audio can leave measurable traces in the frequency domain.

A limitation of many detection settings is that the negative class is often too clean. In a real platform, an uploaded song is rarely only an untouched original or a fully synthetic generation. It may be a remix, a pitch-shifted version, a slowed or sped-up edit, a re-encoded upload, a cover, or a track processed through compression and mastering tools. These edited versions are not AI-generated, but they can introduce spectral and temporal artifacts of their own. A detector evaluated only against clean original songs may therefore learn cues that do not hold in practice. This is especially important for music, where production effects, codec traces, and platform-specific processing can be confused with generation artifacts [Afchar *et al.*, 2025b; Cros Vila *et al.*, 2025].

In this work, we study AI-generated music detection as a hard-negative robustness problem. Instead of focusing on the simpler distinction between AI-generated and original songs, we ask whether a model can distinguish AI-generated music detection from edited or processed variants derived from the same anchor songs. The edited class is treated as a hard negative: it is non-AI content, but it may contain processing traces that can activate AI-like evidence. This framing follows the broader lesson from audio deepfake and spoofing research, where performance under clean conditions does not guarantee robustness in the wild [Liu *et al.*, 2023; Zang *et al.*, 2024]. It also aligns with recent warnings that high AI-music detection scores should be interpreted care-fully when detectors are exposed to audio manipulation or unseen generation pipelines [Afchar *et al.*, 2025b].

To study this problem, we compile a YouTube dataset grouped by reference song comprising AI-generated, edited, and original music variants. A key feature of our implementation is the use of reference song-based splitting, which ensures that all versions derived from the same reference song remain within the same data partition, thereby preventing leakage and ensuring a more rigorous evaluation of model robustness. Our detection pipeline fine-tunes a pretrained PaSST model on 10-second audio clips extracted from each upload.

Our results confirm that edited audio constitutes a significant challenge for AI-music detection. While the system achieves high performance in identifying synthetic content, the lower accuracy on edited samples highlights how common production effects can mimic generative artifacts. To further understand these failure modes, we complement our quantitative findings with a qualitative analysis using Grad-CAM maps to inspect the time-frequency regions that drive high-confidence AI predictions.

The main contributions of this paper are:

- We study edited audio as a hard negative for AI-generated music detection. Since both AI-generated and edited versions differ from the original song, this setting tests whether AI generation leaves spectro-temporal cues that remain distinguishable from ordinary editing and post-processing artifacts.
- We fine-tune a pretrained PaSST spectrogram transformer on 10-second audio clips and evaluate both clip-level and video-level performance.
- We show that edited audio remains a challenging negative class, its lower F1-score indicating confusion with AI-generated audio, and the qualitative analysis suggesting that editing artifacts can overlap with the time-frequency regions used by the detector.

The rest of the paper is organized as follows. Section 2 reviews related work on AI-generated music detection, robustness, and pretrained audio representations. Section 3 describes the dataset and task definition, while Section 4 presents the proposed methodology. Section 5 reports the experimental results, followed by the qualitative analysis and limitations in Sections 6 and 7. Section 8 concludes the paper.

# 2 Related Work

## 2.1 AI-generated music detection

AI-generated music detection is now emerging as a separate research topic, driven by the spread of full-song generators such as Suno and Udio. [Afchar *et al.*, 2025b] showed that audio-based detectors can obtain very high scores in controlled settings, but also stressed that robustness to audio manipulation and generalization to unseen generators remain open problems. [Cros Vila *et al.*, 2025] reached a similar conclusion on a large collection of Suno and Udio tracks, showing that detector performance can be affected by simple transformations such as resampling. These results motivate evaluation settings that go beyond clean AI-versus-original splits. Our work fits into this direction by focusing on a specific robustness case in which edited and processed music is treated as the negative class, rather than only as an external transformation applied after evaluation.

Recent datasets further broaden the scope of this task. SONICS targets end-to-end synthetic song detection using full songs generated with commercial platforms, with emphasis on long-duration structure and song-level classification [Rahman *et al.*, 2024]. FakeMusicCaps focuses on text-to-music detection and attribution across multiple generation models, including closed-set and open-set settings [Comanducci *et al.*, 2025]. Other studies explore complementary signals, such as lyrics transcripts obtained from audio [Frohmann *et al.*, 2025] or symbolic MIDI features for classical music [Tantra and Wicaksana, 2025]. In contrast, our work focuses on waveform audio from online music uploads and studies whether AI-generated variants can be separated from edited variants of related songs.

## 2.2 Artifacts, robustness, and hard negatives

Deepfake detection is widely studied across media modalities [Mirsky and Lee, 2021]. In audio, a common explanation for synthetic-content detection is that generative pipelines leave traces in the signal. [Sun *et al.*, 2023] studied synthetic voice detection through neural vocoder artifacts, while [Afchar *et al.*, 2025a] linked AI-music artifacts to spectral peaks caused by architectural choices such as deconvolution modules. This supports the idea that time–frequency representations can expose useful detection cues.

The difficulty is that non-AI processing can also create artifacts. Online music may be compressed, re-encoded, remixed, equalized, pitch-shifted, time-stretched, or mastered differently from the original. These operations can modify the spectrum and may be mistaken for synthetic evidence. Similar concerns appear in audio deepfake research: ASVspoof benchmarks formalized spoofing evaluation protocols [Wang *et al.*, 2020], ASVspoof 2021 moved spoofing detection toward in-the-wild conditions involving compression, transmission, and data from different sources [Liu *et al.*, 2023], while SingFake showed that singing voice deepfake detection is affected by background music, codecs, languages, singers, and musical context [Zang *et al.*, 2024]. This connects to broader concerns about generalization and robustness in audio deepfake detection [Müller *et al.*, 2022; Wu *et al.*, 2025]. Our work follows this perspective by treating edited audio as a hard negative for AI-generated music detection.

## 2.3 Pretrained audio representations

Large pretrained audio models are now common for downstream audio classification. AudioSet provided the large-scale ontology and labeled data used by many of these models [Gemmeke *et al.*, 2017]. PANNs showed that AudioSet-pretrained networks transfer well to several audio pattern recognition tasks [Kong *et al.*, 2020], while CLAP learned audio representations from natural language supervision and enabled flexible downstream classification [Elizalde *et al.*, 2023].

Spectrogram transformers are especially relevant to our approach. AST introduced a convolution-free transformer for audio spectrogram classification [Gong *et al.*, 2021]. PaSST further improved spectrogram transformers through Patchout, reducing training cost while preserving strong audio classification performance [Koutini *et al.*, 2022]. Following this line of spectrogram-transformer models, we fine tune a pretrained PaSST model on raw audio clips, with the log-Mel representation computed internally by the model frontend.

# 3 Data and Task

## 3.1 Dataset construction

We compiled the dataset from YouTube music uploads starting from a set of reference songs. For each reference song, we searched for related uploads using the artist name, song title, and version-oriented terms such as *AI cover*, *AI version*, *remake*, *remix*, *slowed*, or *edited version*. The goal was to collect different versions of the same underlying song, rather than unrelated examples sampled from a general music catalog.

We used the YouTube Data API to retrieve candidate uploads and metadata. Each entry stores the video identifier, URL, title, channel name, description, artist and song title, together with a reference-song index that links versions of the same underlying song. The reference-song index groups uploads derived from the same original song. Audio was then downloaded with `yt-dlp` and duration checks and decoding were handled with `ffmpeg` and `ffprobe`.

The initial metadata file contained 3,640 entries labeled as *ai*, *edited*, *original*. The final audio subset contains 933 videos from 95 reference-song groups, with 311 AI-generated videos, 311 edited videos, and 311 original videos.

## 3.2 Task formulation

The final classifier is trained on a binary task: *ai* versus *edited*. Original videos are not used as an output class, they are kept as reference material for comparison and prototype-based analysis.

A non-AI upload may still be compressed, remixed, re-encoded, pitch-shifted, time-stretched, or mastered differently from the original recording. These operations can introduce spectro-temporal artifacts that are not caused by AI generation, yet may still look suspicious to a detector. We therefore treat edited audio as a hard negative: it is non-AI content, but it can contain processing traces that overlap with AI-like evidence.

The goal is not to detect whether a song differs from its original version. The goal is to test whether AI-generated versions remain distinguishable from edited versions of related songs.

## 3.3 Grouped split

A random split over clips or videos would be problematic for this dataset. Different uploads can come from the same reference song and may share melody, lyrics, artist identity, structure, or production traits. If one version of a song appeared in training and another version of the same song appeared in testing, the model could learn song-specific cues instead of cues related to AI generation.

To reduce this risk, we split the data by reference-song group. All versions associated with the same reference song are assigned to the same partition. This makes the task harder, but it gives a cleaner estimate of whether the detector generalizes to unseen songs and their related AI or edited variants.

Because the split is performed by reference-song group rather than by individual video, the class counts in each partition are not equal. This is to be expected as not every reference song has the same number of AI, edited, and original uploads. Preventing leakage between related song versions is more important than enforcing a perfectly balanced validation or test set. For this reason, we report balanced accuracy and macro F1-score in addition to accuracy.

Table 1 shows the final split which contains 65 reference-song groups for training, 15 for validation, and 15 for testing. For training, the two target classes were balanced at video level, resulting in 210 AI-generated and 210 edited training videos.

| Split | Groups | AI | Edited | Original | Total |
|---|---|---|---|---|---|
| Train | 65 | 210 | 211 | 235 | 656 |
| Validation | 15 | 56 | 48 | 33 | 137 |
| Test | 15 | 45 | 52 | 43 | 140 |

Table 1: Video-level split by reference-song group.

# 4 Methodology

## 4.1 Audio preprocessing

The model operates directly on audio waveforms rather than on exported spectrogram images. Each available video is decoded as mono audio and resampled to 32 kHz. The audio is then divided into 10-second clips, which are used as the basic input units for training and evaluation. This duration was chosen as a compromise between local artifact detection and enough musical context to capture timbre, production, and short structural patterns.

## 4.2 PaSST-based audio classifier

The main model is based on the pretrained PaSST spectrogram transformer [Koutini *et al.*, 2022]. We use the `hear21passt` implementation with the `passt_s_swa_p16_128_ap476` backbone. The model receives raw waveform clips as input. Its frontend computes the log-Mel representation internally, using 128 Mel bands, an FFT size of 1024, a window length of 800 samples, and a hop size of 320 samples. At 32 kHz, this corresponds to a 10 ms temporal hop.

The resulting log-Mel representation is processed by the pretrained PaSST backbone, and the final classification head predicts one of two classes: *ai* or *edited*. This setup keeps the pipeline closer to an audio classification system than to an image classifier trained on saved spectrogram plots. It also allows the model to benefit from pretrained spectrogram representations while still being fine-tuned on the target task.

### 4.3 Training procedure

Training is performed in two stages. First, the PaSST backbone is frozen and only the classification head is trained. Second, the full model is unfrozen and fine-tuned end-to-end.

The selected configuration uses batch size 4, head learning rate $10^{-3}$, fine-tuning learning rate $8 \times 10^{-6}$, weight decay $8 \times 10^{-5}$, and label smoothing 0.03. We apply light audio augmentation during training: random gain variation up to 1.5 dB and a temporal roll of up to 0.1 seconds. The loss is weighted cross-entropy with label smoothing (0.03); since edited is the harder negative class in our setting, we slightly upweight it during training. Training is run for 2 epochs with the backbone frozen (head-only), followed by up to 6 fine-tuning epochs with early stopping patience of 3.

**Model selection.** We use the validation set for two choices: (i) selecting the best training checkpoint and (ii) selecting the video-level aggregation rule. The checkpoint is chosen by validation balanced accuracy (with macro F1 as a secondary criterion). Aggregation rules (mean, max, count-based, fraction-based) are compared on validation, the best-performing rule is fixed, and results are then reported once on the held-out test set.

### 4.4 Clip and video-level decision rules

The model produces a probability $p(ai)$ for each 10 second clip, i.e., the estimated probability that the clip belongs to the AI-generated class. We use clip-level predictions in two ways. First, for clip-level reporting, the decision threshold is selected on the validation set. This provides a fairer estimate of segment-level precision, recall, and F1-score than fixing the threshold manually.

The final decision is made at video level. A YouTube upload may contain both clear and ambiguous segments, so a single clip is not always representative of the full video. We therefore evaluate candidate aggregation rules on the validation set, including mean, maximum, count-based, and fraction-based aggregation of clip-level AI probabilities.

The selected video-level rule is fraction-based. The best validation rule classified a video as AI-generated if at least 33% of its clips satisfy

$$p(ai) \geq 0.6$$

Otherwise, the video is classified as edited. The 33% fraction was selected empirically on the validation set and reflects the intuition that AI-related traces may not appear uniformly across the whole upload. Some segments may remain ambiguous or resemble ordinary editing; however, if roughly one third of the clips show sufficiently strong AI evidence, the video contains enough consistent signal to be classified as AI-generated. Since this rule was chosen on the validation set, a sensitivity analysis with respect to the fraction and probability thresholds is left for future work.

## 5 Experiments and Results

### 5.1 Experimental setup

All final results are reported on the held-out test split described in Section III-C. The split is performed by reference-song group, so the model is evaluated on songs that do not appear in training or validation. Since the class counts are not perfectly equal after this grouped split, we report balanced accuracy and macro F1-score in addition to standard accuracy.

The PaSST model is trained and evaluated on 10-second audio clips. The final clip-level sets contain 4,447 training clips, 2,389 validation clips, and 2,627 test clips. The test set contains 1,659 AI-generated clips and 968 edited clips. Video-level predictions are obtained by aggregating the clip probabilities for each YouTube upload, using the validation-selected aggregation rule described in the methodology.

### 5.2 Reproducibility and training cost

For reproducibility, all experiments used a fixed global random seed of 42. We performed a limited grid-style hyperparameter search on the validation set; the selected training configuration and the model-selection criteria are described in Section 4.3. Experiments were run on an NVIDIA GeForce RTX 3060 GPU. The full run for the selected configuration, including training and validation aggregation, took approximately 18 hours.

### 5.3 Baselines

Before adopting the final PaSST model, we evaluated two simpler baselines. The first is a Random Forest classifier trained on descriptive audio features. This baseline represents a traditional audio classification approach, where the model uses global descriptors rather than learned spectro-temporal representations. It reached 0.52 accuracy and 0.52 F1-score, thus indicating that handcrafted descriptors alone do not capture enough information for this task.

The second baseline is a MobileNetV2 model trained on exported spectrogram images. This baseline was included to test whether an image-pretrained convolutional network can separate AI-generated and edited audio when spectrograms are treated as images. This model performed better than Random Forest, but remained limited on balanced and macro metrics. As shown in Table 2, these baseline results motivated the move from generic image-based spectrogram classification to a pretrained audio transformer operating directly on waveform input.

| Model | Setting | Acc. | Bal. Acc. | Macro F1 |
|---|---|---|---|---|
| Random Forest | Descr. | 0.520 | – | 0.520 |
| MobileNetV2 | Clip-level | 0.643 | 0.594 | 0.577 |
| MobileNetV2 | Video-level | 0.739 | 0.601 | 0.598 |

Table 2: Baseline results. RF uses audio descriptors; MobileNetV2 uses spectrograms.

### 5.4 Clip-level results

Table 3 reports the final PaSST results at clip level on the held-out test set. The stronger performance on the AI class suggests that the model captures useful generation-related cues, while the lower edited-class score confirms that edited audio is a difficult negative class.

The gap between the two classes is important for the goal of this paper. Edited clips are not clean negatives. They may contain compression, re-encoding, pitch or tempo changes, and other processing traces. Some of these traces can overlap with the spectro-temporal cues used by the detector. This explains why the edited class is harder, and it supports the hard-negative framing of the task.

| Class | Precision | Recall | F1 | Support |
|---|---|---|---|---|
| AI-generated | 0.8366 | 0.8360 | 0.8363 | 1659 |
| Edited | 0.7193 | 0.7200 | 0.7197 | 968 |
| Macro avg. | 0.7780 | 0.7780 | 0.7780 | 2627 |

Table 3: Clip-level PaSST results on the held-out test set.

### 5.5 Video-level results

At deployment time, the system would usually need to assign one label to a full uploaded video, not to an isolated 10 second segment. We therefore aggregate the clip predictions for each upload and report video-level results. This evaluation reflects the intended use case, since a video may contain both clear and ambiguous segments.

The video level scores in Table 4 are higher than the clip-level macro F1-score, which suggests that aggregation helps reduce the effect of ambiguous or noisy clips. A single suspicious segment is not enough to classify the whole video as AI-generated, the final decision depends on the distribution of AI evidence across the upload.

| Metric | Score |
|---|---|
| Accuracy | 0.8173 |
| Balanced accuracy | 0.8110 |
| Macro F1 | 0.8134 |
| AI-generated F1 | 0.8403 |

Table 4: Video-level PaSST results on the held-out test set.

### 5.6 Result interpretation

The results support three observations. First, the task is not solved by simple audio descriptors. The Random Forest baseline remains close to chance, suggesting that global descriptors miss many of the local spectro-temporal patterns needed for this distinction. Second, treating spectrograms as images with MobileNetV2 improves over handcrafted features, but it does not provide strong balanced performance. This is expected, since the model is pretrained for natural images rather than audio structure.

Third, the PaSST model provides a clear improvement and reaches useful video-level performance under the edited-audio hard-negative setting. These results should be read as evidence for this specific hard-negative setting, not as a general-purpose detector for all AI-generated music. It shows that pretrained spectrogram transformers can separate AI-generated and edited music variants to a useful degree, while edited audio remains a major source of confusion. For a robustness-oriented evaluation, this remaining confusion is part of the finding: non-AI processing can produce evidence that overlaps with AI-like spectral cues.

## 6 Qualitative Analysis

We use the qualitative analysis to inspect the model behavior in the time-frequency domain. The log-Mel spectrograms show how AI-generated and edited clips from the same reference song differ in energy distribution and harmonic structure, while Grad-CAM maps indicate which regions of the spectrogram influence the model's predictions. These visualizations help us to understand how to interpret the model behavior without claiming a universal AI fingerprint.

### 6.1 Reading the log-Mel spectrogram

Given a waveform $x[n]$, we first compute a short-time Fourier transform (STFT):

$$X(t,k) = \sum_{n=0}^{N-1} x[n+tH]w[n]e^{-j2\pi kn/N},$$

where $t$ is the time frame, $k$ is the frequency bin, $w[n]$ is the analysis window, $N$ is the FFT size, and $H$ is the hop length. The power spectrum is then projected onto a Mel filterbank:

$$E(t,m) = \sum_{k} M_{m,k}|X(t,k)|^2,$$

where $M_{m,k}$ is the weight of Mel filter $m$ at frequency bin $k$. The displayed representation is the log-Mel energy:

$$S(t,m) = \log(E(t,m) + \epsilon),$$

with a small $\epsilon$ added for numerical stability.

In Figure 1, the horizontal axis represents time frames, with a 10 ms hop between adjacent frames. The vertical axis represents frequency in kHz. Brighter colors indicate higher log-Mel energy, while darker regions indicate lower energy. Horizontal bands usually correspond to harmonic musical content, such as vocals or pitched instruments. Vertical structures often indicate onsets, transients, abrupt edits, or other short-time changes in the signal.

The AI-generated and edited clips in Figure 1 come from the same reference song. This makes the comparison more meaningful, since both clips share musical content. The AI-generated clip shows strong repeated harmonic bands and dense low-frequency energy. The edited clip preserves related musical structure, but its energy distribution differs, especially in the mid and high frequency ranges. This illustrates the difficulty of the task, edited audio is not a clean negative class, since post-processing can also create visible time frequency artifacts.

### 6.2 Grad-CAM for spectrogram-based decisions

To inspect which regions support a class prediction, we use Grad-CAM on the spectrogram representation. For a target class $c$, let $A^k$ be activation map $k$ from the selected internal layer of the model, and let $y^c$ be the class score before the final softmax. Grad-CAM computes an importance weight for each activation map:

$$\alpha_k^c = \frac{1}{Z}\sum_i \sum_j \frac{\partial y^c}{\partial A_{ij}^k},$$

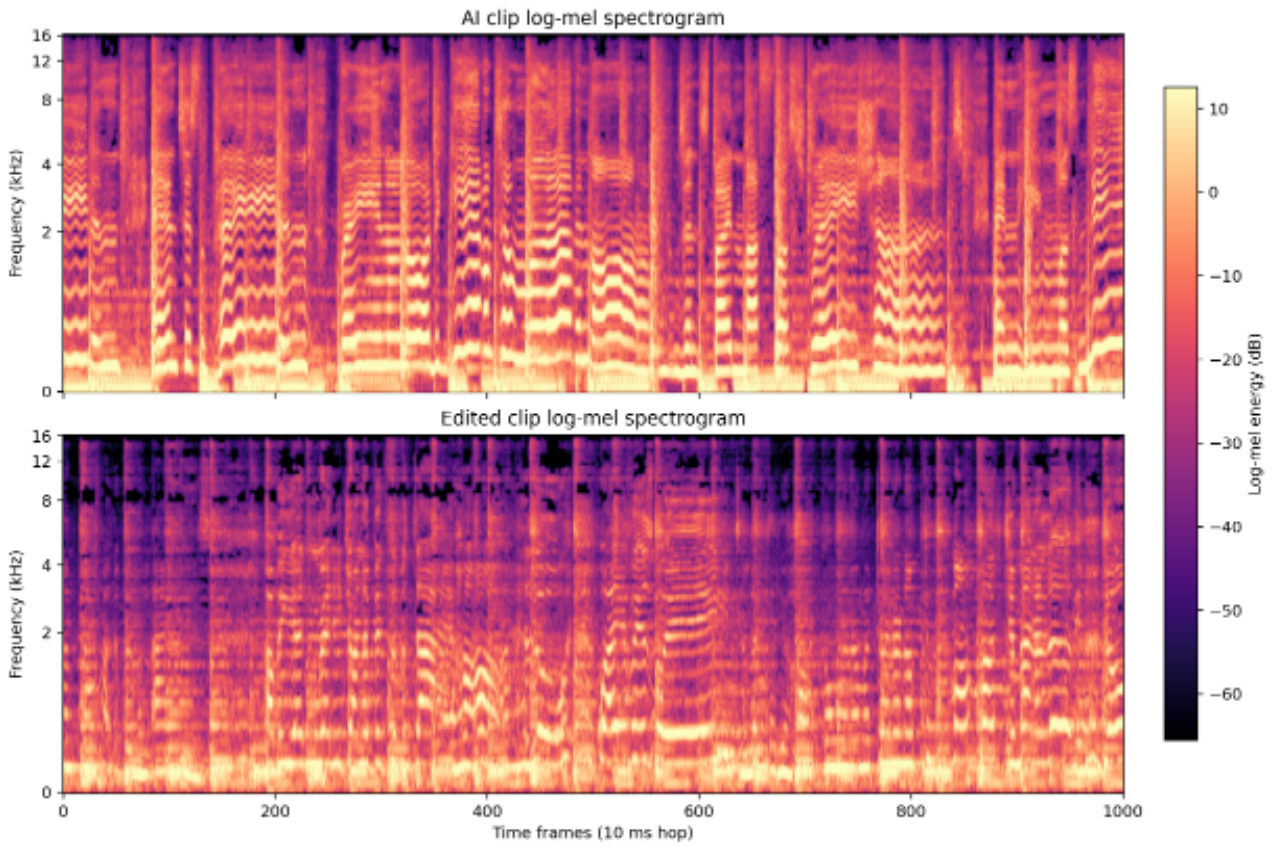

Figure 1: Log-Mel spectrogram comparison for an AI-generated clip and an edited clip from the same reference song.

where $Z$ is the number of spatial locations in the activation map. The class activation map is then computed as

$$L^{c}_{\mathrm{Grad-CAM}} = \mathrm{ReLU}\left(\sum_{k} \alpha^{c}_{k} A^{k}\right).$$

The resulting map is upsampled to the spectrogram resolution and normalized between 0 and 1. Warmer colors indicate time–frequency regions that contribute more strongly to the selected class prediction. In this setting, a high Grad-CAM value does not mean that a region is uniquely AI-generated or uniquely edited. It means that the region was important for the model's current decision.

Figure 2 shows Grad-CAM maps for a same-song AI/edited pair. The AI clip is correctly classified as AI-generated with $p(ai) = 0.909$. The corresponding Grad-CAM map highlights localized regions around harmonic and transient-like structures, rather than uniformly activating over the full spectrogram. This suggests that the model is not relying only on global loudness or average spectral energy.

The edited clip is correctly classified as edited with $p(edited) = 0.887$. Its Grad-CAM map highlights a different set of time–frequency regions, including localized activations in the mid and upper Mel bins. These regions are consistent with the idea that post-processing and editing can introduce localized spectral changes. At the same time, the two maps are not cleanly separable into a simple "AI area" and "edited area". Both examples contain structured activations, which explains why edited audio remains a hard negative.

These visualizations support the interpretation of the numerical results. The model uses localized spectro-temporal evidence, and this evidence differs between the AI-generated and edited examples. The remaining overlap between the activation patterns is also important: it matches the lower edited-class F1-score and reinforces the main argument that edited audio should be treated as a hard negative when evaluating AI-generated music detectors.

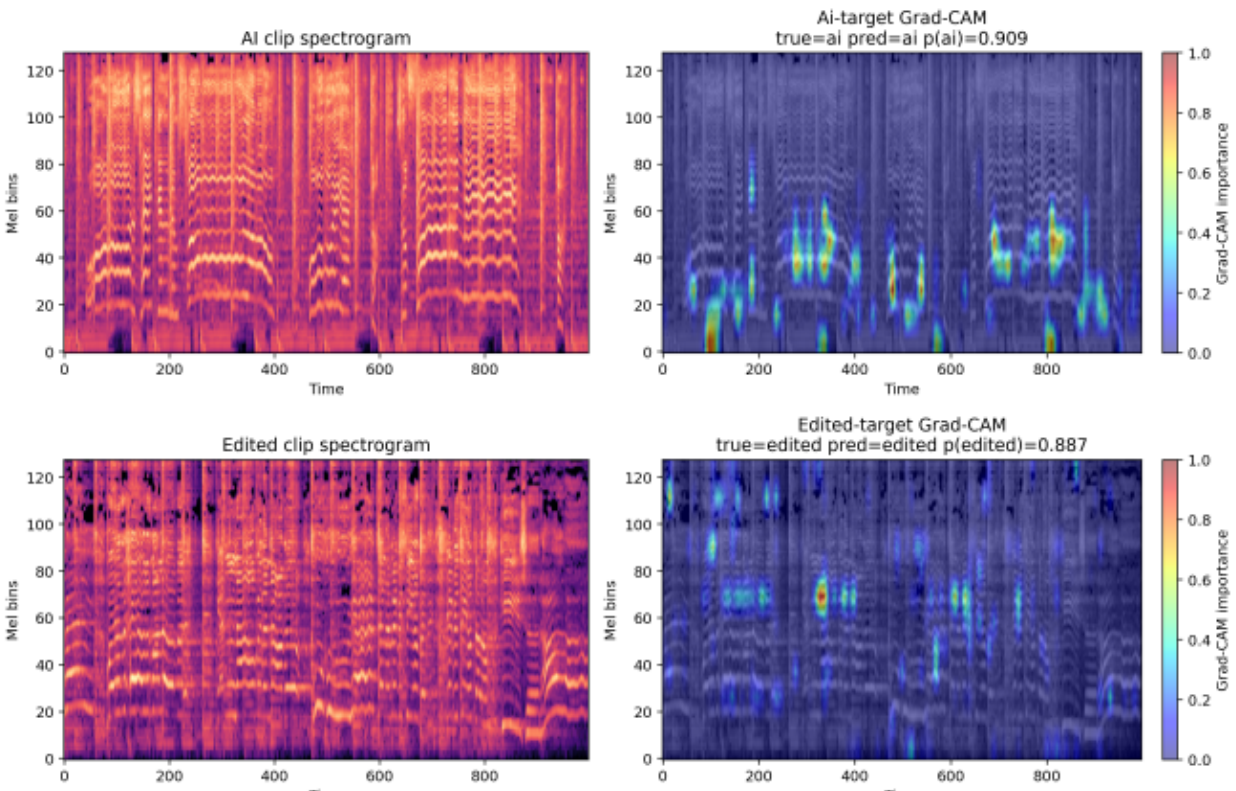

Figure 2: Grad-CAM visualization for the same-song AI/edited pair.

## 7 Limitations

This study focuses on AI-generated and edited music variants collected from online uploads. From the upload alone, it is not always possible to know whether a track was only edited by a human, altered with AI-assisted tools, or processed through a combination of both. For this reason, the edited class should be interpreted as a non-AI or non-fully-AI hard negative based on the available metadata and manual labeling, rather than as a perfectly controlled editing category.

The dataset also does not cover all musical styles equally. Our collection is centered on the reference songs and online variants available through the search procedure, so some genres, such as classical music or purely instrumental music, are underrepresented. Since different genres have different spectral structure, production practices, and dynamic range, performance may vary on music styles outside the current dataset.

The analysis is based on naturally occurring edited uploads, not on a controlled set of individual transformations. This means that the model is evaluated in a realistic setting, but the results do not isolate the effect of each editing operation separately. Future work could extend the dataset with controlled transformations such as compression, pitch shifting, time stretching, and re-encoding to measure which types of processing most often produce AI-like evidence.

Finally, the empirical evaluation remains limited in baseline coverage. While we include two lightweight reference baselines (Random Forest on handcrafted audio descriptors and MobileNetV2 on exported spectrogram images), we do not compare against stronger pretrained audio representation models such as CLAP, PANNs, AST-family variants, or AudioMAE-style approaches, nor against recent AI-audio detection methods. Extending the baseline suite and evaluating cross-model and cross-platform generalization are important directions for future work.

## 8 Conclusion

This paper investigates AI-generated music detection under a hard-negative setting, where the negative class is not clean original music but edited and processed audio. This setting is closer to a practical failure case for online music platforms: a non-AI upload may still contain compression, remixing, pitch or tempo changes, re-encoding, or other processing traces that can look suspicious to an automatic detector.

The main contribution of this work is the exploration of edited audio as a hard negative in AI-generated music detection, showing that non-AI processing can introduce artifacts that overlap with AI-like spectral evidence. We compiled a YouTube-based dataset of related AI-generated, edited, and original song versions, and used reference song splitting to reduce leakage between different versions of the same song. Original recordings are kept as reference material, while the final classifier is evaluated on the more difficult distinction between AI-generated and edited audio. This setup tests whether a detector can separate AI evidence from non-AI processing evidence, rather than simply detecting whether a song differs from a clean original.

Using audio clips and a pretrained PaSST spectrogram transformer, the final video-level system reaches 0.817 accuracy, 0.811 balanced accuracy and 0.813 macro F1-score on the held-out test set. These results show that pretrained spectrogram transformers can capture useful cues for separating AI-generated and edited audio. At the same time, the lower edited-class performance at clip level shows that edited audio is not an easy negative class. It can contain processing traces that overlap with the evidence used for AI detection.

The conclusion is not that AI-generated music detection is solved. Rather, our results suggest that evaluating detectors only against clean original songs is too optimistic. Edited audio should be included as a robustness test, since it exposes a realistic source of false AI evidence. Future work should extend this evaluation to more generators, more editing operations, and larger cross-platform datasets, while also studying how detector decisions change under controlled audio transformations.

## Acknowledgment

This research is supported by the project "Romanian Hub for Artificial Intelligence - HRIA", Smart Growth, Digitization and Financial Instruments Program, 2021-2027, MySMIS no. 351416.